\documentclass[useAMS]{mn2e}
\usepackage{psfig}
 \usepackage{url}
\usepackage{times}
\def\spose#1{\hbox to 0pt{#1\hss}}
\newcommand\lsim{\mathrel{\spose{\lower 3pt\hbox{$\mathchar"218$}}
     \raise 2.0pt\hbox{$\mathchar"13C$}}}
\newcommand\gsim{\mathrel{\spose{\lower 3pt\hbox{$\mathchar"218$}}
     \raise 2.0pt\hbox{$\mathchar"13E$}}}
\def\sc{Schwarzschild}
\def\ltsima{$\; \buildrel < \over \sim \;$}
\def\lsim{\lower.5ex\hbox{\ltsima}}
\def\gtsima{$\; \buildrel > \over \sim \;$}
\def\gsim{\lower.5ex\hbox{\gtsima}}

\title[Dark bubbles around high--$z$ radio--loud AGN]
{Dark bubbles around high--redshift radio--loud AGN}
\author[G. Ghisellini et al.]
{G. Ghisellini$^1$, \thanks{E--mail: gabriele.ghisellini@brera.inaf.it}
T.  Sbarrato$^{2}$ \\ \\
$^1$ INAF -- Osservatorio Astronomico di Brera, via E. Bianchi 46, I--23807 Merate, Italy \\
$^2$ Univ. di Milano Bicocca, Dip. di Fisica G. Occhialini, Piazza della Scienza 3, I--20126 Milano, Italy \\
}

\begin{document}

\pagerange{\pageref{firstpage}--\pageref{lastpage}} \pubyear{2012}

\maketitle
\label{firstpage}

\begin{abstract}
At redshift larger than 3 there is a disagreement 
between the number of blazars (whose jet is pointing at us) and the number of expected parents 
(whose jet is pointing elsewhere).
Now we strengthen this claim because (i) the number of blazars 
identified within the SDSS+FIRST survey footprint increased, 
demanding a more numerous parent population,  and (ii) the detected blazars have
a radio flux large enough to be above the FIRST flux limit even if the jet is slightly misaligned.
The foreseen number of these slightly misaligned jets, in principle detectable, is much larger than
the radio--detected sources in the FIRST+SDSS survey (at redshift larger than 4).
This argument is independent of the presence of an isotropic radio component,
such as the hot spot or the radio lobe, and does not depend on the bulk Lorentz factor $\Gamma$.
We propose a scenario that ascribes the lack of slightly misaligned sources to an over--obscuration 
of the nucleus by a ``bubble" of dust, possibly typical of the first high--redshift quasars.
\end{abstract}

\section{Introduction}

Blazars (flat spectrum radio quasars, FSRQs, and BL Lac objects) produce most of
their non--thermal radiation in jets whose plasma is moving relativistically
at small angles $\theta$ from the line of sight. 
How small 
the viewing angle must be 
for a source to be a blazar is not defined exactly, but we have proposed to use $\theta< 1/\Gamma$,
where $\Gamma$ is the bulk Lorentz factor of the emitting plasma.
Under this definition, for a jet  pointing in our direction within an angle $1/\Gamma$  there 
must exist $2\Gamma^2$ other  sources
with their jets pointing elsewhere: these sources form the parent population of blazars,
and are usually associated with the FR I (low luminosity) and FR II (high luminosity)
radio--galaxies (Fanaroff \& Riley 1974). 

Volonteri et al. (2011; hereafter V11) pointed out the difficulties in reconciling the 
number of blazars observed at high redshifts with the number of the expected parent
population.
The flux of these sources is less beamed and amplified with respect to the 
aligned sources, and for large enough $\theta$ is even de--beamed, but 
the extended structures at the end of the jets (hot spots and lobes),
that emit isotropically, could be bright enough to be detectable,
especially if the jet is powerful (FR II type), as are all the
jets detected at high redshifts.

V11 also pointed out that the disagreement between the number of blazars and 
their parents occurs only for redshifts $z\gsim 3$.
This was based on two cross correlated catalogs: the Fifth Quasar Catalog (Schneider et al.\ 2010)
of the Sloan Digital Sky Survey (SDSS; York et al.\ 2000) and the 
Faint Images of the Radio Sky at Twenty--Centimeters
(FIRST; White et al.\ 1997).
The SDSS Quasar Catalog is a spectroscopic, magnitude limited quasar survey 
($m_{i}<19$ or 21 for low-- and high--redshift quasars)
and FIRST is a VLA radio survey complete above 1 mJy at 1.4 GHz.
The common sky area is 8,770 square degrees.
The quasars belonging to the two samples (that we will collectively call SDSS+FIRST) 
have been studied by Shen et al. (2011).
According to this study, and listed in Tab. \ref{numbers},
in the redshift bin $4<z<5$ there are 1192 quasars, 49 of which are radio-detected.
Of these, at least 6 are blazars.
Above redshift 5 there are 56 quasars, of which 4 are radio--detected,
and 2 of these are blazars.

If we take the 6 blazars with redshift between 4 and 5, 
we expect $(1200\pm 480)(\Gamma/10)^2$ misaligned jets
(assuming an uncertainty of $\sqrt{6}$ in the number of observed 
blazars in this redshift bin),
but we see a total of only 49 radio sources above 1 mJy.
Above $z>5$, the 2 observed blazars 
should correspond to $(400\pm280)(\Gamma/10)^2$ parents,
but we see a total of only 4 sources above 1 mJy.
These numbers strengthen the problem pointed out in V11, because, since then, more radio sources
in the SDSS+FIRST turned out to be blazars (see e.g. Sbarrato et al. 2013; 
Ghisellini et al. 2014; Sbarrato et al. 2015).

V11 proposed three possible solutions to this disagreement:
i) the bulk Lorenz factor is much lower than what it is at $z\lsim 3$.
To reconcile the numbers, we would need $\Gamma=2$, which is inconsistent with the
observed properties of high--$z$ blazars 
(see, e.g. Ghisellini \& Tavecchio 2015; Lister et al. 2013); 
ii) there is a (yet unknown) bias in the SDSS+FIRST survey against the detection of
high--$z$ radio--loud sources. 
For instance, the isotropic radio structure could be young and compact,
self--absorbed at frequencies larger than 10 GHz (in the rest frame), and therefore be
below the 1 mJy flux detection limit of the FIRST;
iii) the SDSS+FIRST survey misses the detection of a large population of parents because their
optical flux is absorbed by dust.
A fourth solution would be to postulate the absence of the hot spot and lobe in these
sources, or that these structures are very faint in the radio band for redshifts larger than 3.

In fact, the $\propto(1+z)^4$ scaling of the cosmic background radiation (CMB) energy density
can greatly affect the radio emission of extended structures, as explored by 
Mocz, Fabian \& Blundell (2011) and by Ghisellini et al. (2013; 2015).
Consider two sources at different redshifts, that have the same size, magnetic field, and that
are energized by the same injected power.
The higher--$z$ source will have a fainter radio emission and a stronger X--ray luminosity
than the lower--$z$ source.
This is because the emitting electrons will preferentially cool through  
inverse Compton scattering off CMB seed photons rather than producing synchrotron emission.
The quenching of the radio emission can help to reconcile the disagreement between 
the number of high--$z$ blazars and the corresponding number of expected parents, 
but in this paper we point out that this effect is not enough.

In fact we will point out that also the sources whose jet is slightly misaligned
can produce a flux above the threshold limit of the radio survey.
This is independent of the CMB energy density.
If a jet emits a radio flux of -- say -- 200 mJy and it is observed at a given (small) 
$\theta$, there should be other similar jets observed at larger viewing angles but whose 
radio flux is larger than the flux limit of the survey (in our case, 1 mJy).
Even these sources are missing. 
The problem is even more severe because we will show that 
the expected number of these sources is independent of the bulk Lorentz factor.
We believe that this calls for a revision of our basic understanding of
these high--$z$ sources.

\section{Slightly misaligned jets}

We define as blazar a source whose jet is observed at a viewing angle $\theta\le 1/\Gamma$.
At $\theta=1/\Gamma$, the Doppler factor is
\begin{equation}
\delta\equiv \frac{1}{\Gamma(1-\beta\cos\theta)} = \Gamma.
\end{equation}
Smaller angles have larger Doppler factors (at $0^\circ$, $\delta\sim 2\Gamma$), but the probability $P$
to observe a jet pointing exactly at us is vanishingly small ($P\propto \theta^2$).
Assume that a source, in the comoving (primed) frame, emits a 
monochromatic flux $F^\prime(\nu^\prime)=F^\prime(\nu/\delta)$.
Then the observer at Earth will see a flux $F(\nu)$: 
\begin{equation}
F(\nu) \, =\, \delta^{p} F^\prime(\nu) 
\end{equation}
The exponent $p$ can have different values.
If the emission is a power law of spectral index $\alpha$ [i.e.\ $F(\nu)\propto \nu^{-\alpha}$]
we have, among the several possibilities:
%

\begin{table} 
\centering
\begin{tabular}{l l r r }
\hline
\hline
SDSS Name &$z$   &$R$   &$F_R$  \\
\hline   
083946.22 +511202.8	 &4.390   &285       &41.6   \\
102623.61 +254259.5  &5.304   &5200      &239.4 \\ 
114657.79 +403708.6  &5.005   &1700      &12.5   \\
130940.70 +573309.9  &4.268   &133       &11.3 \\
132512.49 +112329.7  &4.412  &879        &71.1 \\
142048.01 +120545.9  &4.034  &1904       &87.3 \\ 
151002.92 +570243.3  &4.309   &13000     &255.0   \\
222032.50 +002537.5  &4.205   &4521      &116.0  \\
\hline
{\it 143023.7  ~~+420436} &{\it 4.715}    &{\it 5865}     &{\it 215.6} \\  
{\it 171521.25 +214531.8}  &{\it 4.011}   &{\it 30000}    &{\it 396.0} \\ 
{\it 213412.01 --041909.9} &{\it 4.346}   &{\it 24000}    &{\it 295.1}  \\
\hline
\hline
\end{tabular}
\caption{
The top part of the table lists the known blazars at $z\ge4$ in the SDSS+FIRST
spectroscopic catalog (Shen et al. 2011).
The bottom part lists (in italic) other 3 blazars present in the photometric catalog
of the SDSS+FIRST, but not in the spectroscopic one. 
They are shown for completeness, but we ignore them in the following.
The radio--loudness 
$R$ is defined as $F_{5 {\rm GHz}}/F_{2500{\rm \AA}}$, 
where $F_{5 {\rm GHz}}$ and $F_{2500{\rm \, \AA}}$ are the monochromatic 
rest frame fluxes at 5 GHz and at 2500 \AA, respectively.
$F_R$ is the radio flux density at 1.4 GHz in mJy.
}
\label{sample}
\end{table}

\begin{itemize}
\item $p=2+\alpha$: in this case the jet emits between two locations that are 
stationary in the observer frame. 
The radiation is emitted isotropically in the comoving frame.
Sometimes this is called the finite lifetime jet case.

\item $p=3+\alpha$: this is the case of a moving blob, 
emitting isotropically in the comoving frame. 

\item $p=4+2\alpha$: the jet is a moving blob emitting inverse Compton radiation
using seed photons that are produced externally to the jet 
(the so called external Compton mechanism), and that are
distributed isotropically in the observer frame.
The inverse Compton flux is not isotropic in the comoving frame, 
but it is enhanced in the forward direction (see Dermer 1995). 
This changes the pattern of the radiation as seen in the observer frame with respect to
the previous ($p= 3+\alpha$) case. 
\end{itemize}

\begin{table*} 
\centering
\begin{tabular}{  r | r c c  c | c c   }
\hline
~  &\# Total &\# Radio Det.  &\# Blazars &Obs. ratio  &$R_{\rm tot}(p=2)$ &$R_{\rm tot}(p=3)$    \\
\hline
$4\le z< 5$ &1192 &49 &6 &8.2  &102.7 &45   \\
$z\ge 5$    &56   &4  &2 &2    &36  &15  \\
\hline
\end{tabular}
\caption{
Numbers of radio--detected quasars in the SDSS+FIRST spectroscopic sample and
number of predicted misaligned objects.
We have applied Eq. \ref{eq7} for each blazars considering its actual radio flux
and the limiting flux of 1 mJy of the FIRST survey. 
For each blazar, there are $2\Gamma^2 \sim 338 (\Gamma/13)^2$
jets pointing in other directions.
The number of blazars refers to the objects spectroscopically observed for the
construction of the catalog. There are other high--$z$ blazars in the SDSS+FIRST sky area,
that were photometrically detected, but that were not followed up spectroscopically
(see Table \ref{sample} and Ghisellini et al. 2015).
}
\label{numbers}
\end{table*}

%
Assuming that the maximum observable boost is for 
$\sin\theta\sim 1/\Gamma \rightarrow \cos\theta=\beta$:
\begin{equation}
F_{\rm max}(\nu)\, = \, \delta_{\rm max}^{p}  F^\prime(\nu )\, 
=\,  F^\prime(\nu )  \Gamma^{p}
\end{equation}
The maximum viewing angle $\theta_{\rm c}$ at which this source can be seen is set by
\begin{equation}
F_{\rm min}(\nu) \, =\, \delta_{\rm  min}^{p}  F^\prime(\nu ) \, =\, 
{ F^\prime(\nu) \over \left[\Gamma( 1-\beta\cos\theta_{\rm c})\right]^{p} }
\end{equation}
Therefore the ratio of the maximum to the minimum fluxes gives:
\begin{equation}
{F_{\rm max}(\nu)\over  F_{\rm min}(\nu)} \, =\, 
\left[ {\delta_{\rm max} \over \delta_{\rm min} } \right]^{p} \,
=\, \left[  \Gamma^2 (1-\beta\cos\theta_{\rm c}) \right]^{p} 
\end{equation}
This gives the maximum viewing angle as:
\begin{equation}
\cos\theta_{\rm c} \, =\, {1\over \beta} - \, {1\over \beta\Gamma^2}\, 
\left[{F_{\rm max}(\nu)\over  F_{\rm min}(\nu)}\right]^{1/p}, \quad \beta>0
\end{equation}
Now we can calculate the ratio of the number of sources oriented within $\theta_{\rm c}$ to
the sources oriented within $\theta=1/\Gamma$:
\begin{eqnarray}
R &\equiv & {\rm \#\, within \, \theta_{\rm c} \over \#\, within \, 1/\Gamma}   \, =\, 
{ \int_{\cos\theta_{\rm c}}^1 d\cos\theta \over \int_{\beta}^1 d\cos\theta }\, 
=\, {1-\cos\theta_{\rm c} \over 1-\beta}
\nonumber \\
&=& (1+\beta)\Gamma^2 \left\{  1- { 1\over \beta} + \, {1\over \beta\Gamma^2}
\left[{F_{\rm max}(\nu)\over  F_{\rm min}(\nu)}\right]^{1/p}\right\}
\nonumber \\
&\sim & 2 \left[{F_{\rm max}(\nu)\over  F_{\rm min}(\nu)}\right]^{1/p}\, -1
\label{eq6}
\end{eqnarray}
where the last approximate equality is valid for $\beta\to 1$.
{\it
In this limit the ratio $R$ is not dependent on $\Gamma$.}

As an example, assume that the brightest radio source in a sample is a blazar with 
$F_{\rm max}(\nu)=100$ mJy and the limiting flux of the same sample is 1 mJy. 
Assume $\alpha=0$ and $p=3$.
The very existence of this blazar implies the existence of other 
$[2\times 100^{1/3}-1] = 8.3$ observed {\it jets}, independent of the 
presence or absence of isotropic extended hot spots/lobes.
The limiting viewing angle does depend on $\Gamma$, and is
 $\theta_{\rm c}\sim 17.5^\circ$ (if $\Gamma=10$) or 8.7$^\circ$
(if $\Gamma=20$).

\section{Predicted vs observed radio--loud sources}

We can apply the calculations detailed in the previous section to 
all the blazars in a sample.
Adopting for each of them its radio flux $F_{\rm i}(\nu)$ and the same flux limit (i.e. 1 mJy)
we can calculate for each blazar how many slightly misaligned jets
we expect to be observable, and then sum up to obtain the total expected
ratio between the total detectable sources and the total observed blazars:
\begin{equation}
{R_{\rm tot}} \, =\, \sum_{i=1}^{i=6} 2 \left[{F_{\rm i}(\nu)\over  1 \, {\rm mJy} } \right]^{1/p}\, -1
\label{eq7}
\end{equation}

In few previous works we classified a number of blazar candidates 
included in the SDSS+FIRST 
(Ghisellini et al.\ 2014; Sbarrato et al.\ 2012; 2013; 2015).  
To identify the most reliable candidates, we selected the $z>4$ quasars with a
radio--loudness $R=F_{\rm 5GHz}/F_{\rm 2500\, \AA}>100$ from SDSS+FIRST. 
In fact, among radio--loud sources, a more extreme radio--to--optical dominance 
is a first indication of a jet oriented roughly towards us. 
X--ray follow--up observations allowed us to confirm the blazar classification 
of the most radio--loud candidates (and more observations will follow). 
From these studies, we concluded that there are (at least) 6 blazars 
in the SDSS+FIRST spectroscopic catalog at $4\le z<5$, and 2 at $z>5$. 
These are listed in the top part of Tab. \ref{sample}.
Note that the three sources included in the SDSS+FIRST, but not in its spectroscopic catalog 
(classified in Ghisellini et al.\ 2015) are here excluded because they 
do not have the necessary optical flux to enter the SDSS+FIRST 
spectroscopic, flux--limited sample.
They are listed in the bottom part of Tab. \ref{sample}. 

We  are confident that these blazars are indeed observed at an angle $\theta<1/\Gamma$.
In fact the X--ray flux and spectrum (due to external Compton)
are dependent on the viewing angle in a stronger way
than the radio (synchrotron) flux, as mentioned in \S 2 (the $p$--value is different).
At high redshift (and in the absence of a detection in the $\gamma$--ray band), 
this is the best diagnostic to derive the jet orientation
(see Fig. 3 of Sbarrato et al. 2015 showing how the SED changes by small changes in the
viewing angle).
Furthermore, in the case of B2 1023+25, the small viewing angle and large Lorentz factor
were confirmed  by the european VLBI (EVN) observations by Frey et al. (2015).


The existence of these blazars, compared with the whole SDSS+FIRST radio--detected sample, 
highlights a large discrepancy regarding the number of slightly misaligned jets. 
As listed in Table \ref{numbers}, the radio fluxes of the 
6 blazars 
allow to predict a total of 616$\pm$246 [$(6\pm \sqrt{6})\times R_{\rm tot}(p=2)$] 
or 270$\pm$108 [($6\pm \sqrt{6})\times R_{\rm tot}(p=3)$] jetted sources detectable in the 
SDSS+FIRST survey in the $4\le z<5$ redshift bin.
At $z>5$, this number is 72$\pm$50 ($p=2$) or 30$\pm$21 ($p=3$).
We believe that this is a severe disagreement, because:

\begin{enumerate}

\item the number of expected slightly misaligned objects derived from the known blazars 
	is robust, because independent of $\Gamma$.
	
\item Since the flux comes from the jet, these objects are observed as point--like sources. 
This bypasses the problem of associating one (or two, in the case of a double radio source) 
radio objects not coincident with a SDSS source. 
Furthermore, point--like sources are easier to detect with respect to extended ones.

\item All high--$z$ blazars have their optical flux completely dominated 
by the accretion disc radiation. 
The synchrotron emission (that can be depressed more than the radio in slightly 
misaligned sources, since $\alpha\sim$1) does not contribute significantly to the optical flux. 
This implies that slightly misaligned sources should in principle be included in the 
SDSS quasar catalog.

\item The presence of a dusty torus should not affect the optical flux, 
as long as its opening angle is similar to lower redshift sources. 
This implies that the optical emission, in a standard scenario, should not be obscured.

\end{enumerate}

\section{Obscuring bubbles: a way out}

The discrepancy between the predicted and observed number of sources that have
slightly misaligned jets is serious, and calls for an explanation.
In addition, we are not aware of any instrumental selection effect strongly biasing our sample.
The possibility that were proposed previously by V11 aimed to account for the
lack of extended and isotropically radio sources, namely the foreseen 
parent population of high--$z$ blazars.
To explain these (still missing) sources we can envisage two possible reasons:
i) the observational difficulty to detect a weak extended radio source at some angular distance from a
point--like optical object and ii) the ``radio quenching" effect due to the enhanced CMB radiation
energy density that cools more efficiently the emitting electrons through the inverse Compton mechanisms
and that weakens their radio emission. 

However, the regions of the jet producing the 1.4 GHz radio flux ($\gsim$ 7 GHz rest frame)
are not affected by the ``quenching" of the radio due to the CMB radiation.
This is because they have a magnetic energy density much larger than the CMB one, 
even taking into account the $\Gamma^2$ enhancement due to the relativistic 
motion of the emitting plasma.

At $z=4$, the CMB energy density is $U_{\rm CMB}\sim 2.6\times 10^{-10}$ erg cm$^{-3}$.
In the comoving frame, this is enhanced by a factor $\sim \Gamma^2$, thus reaching
$U^\prime = 2.6\times 10^{-8}(\Gamma/10)^2$ erg cm$^{-3}$.
Most of the observed radiation from the jet is produced in a compact region, where the magnetic field
is around 3 G, and the observed self--absorption frequency is $\nu_{\rm t}\sim 3\times 10^{12}$ Hz
(rest frame, see e.g. Ghisellini \& Tavecchio 2015).
In the case of a flat radio spectrum, the self--absorption frequency scales as $R^{-1}$,
where $R$ is the distance from the black hole.
This is the same dependence of the dominant component of the magnetic field $B$.  
Therefore in the region self--absorbing at 7 GHz (rest frame) $B$
should be $\sim 6$ mG, and its energy density $U_{\rm B}=B^2/(8\pi)\sim 1.4\times 10^{-6}$.
Since $U_{\rm B}>U^\prime_{\rm CMB}$, there is no ``quenching" of the 
synchrotron emission of the jet.

\subsection{The proposed scenario}

To solve the tension between predicted and observed sources,
we propose a scenario that follows the ideas put forward by Fabian (1999).
At redshifts larger than $\sim$4, jetted sources hosting a black hole with mass $M\gsim 10^9 M_\odot$
\footnote{only the quasars with very massive black holes can be detected in the SDSS}
are completely (i.e. 4$\pi$) surrounded by obscuring material. 
Only the jet can pierce through this material and break out.
Observers looking down the jet can see the nuclear emission from the
accretion disc and the broad emission lines.
For observers looking with viewing angles even only slightly larger than 
$\theta_{\rm j}\approx 1/\Gamma$,
the optical emission (including broad lines) is absorbed, the flux is fainter,
and the source cannot enter the SDSS catalog.
The absorbed radiation is re--radiated in the infrared.

Even if the hot spots or the lobes were indeed emitting a radio flux above the 1 mJy level 
at the observed 1.4 GHz frequency, there would be no source in the SDSS to match with
if the jet is only slightly misaligned.
The quasi--spherical dusty structure (hereafter ``obscuring" or ``dark bubble") can cover
the nuclear region until the accretion disc radiation pressure blows it away.
This can occur at a threshold luminosity $L_{\rm th} =\eta_{\rm d}\dot M c^2$.

%

\begin{figure} 
\vskip 0.2 cm
\psfig{file=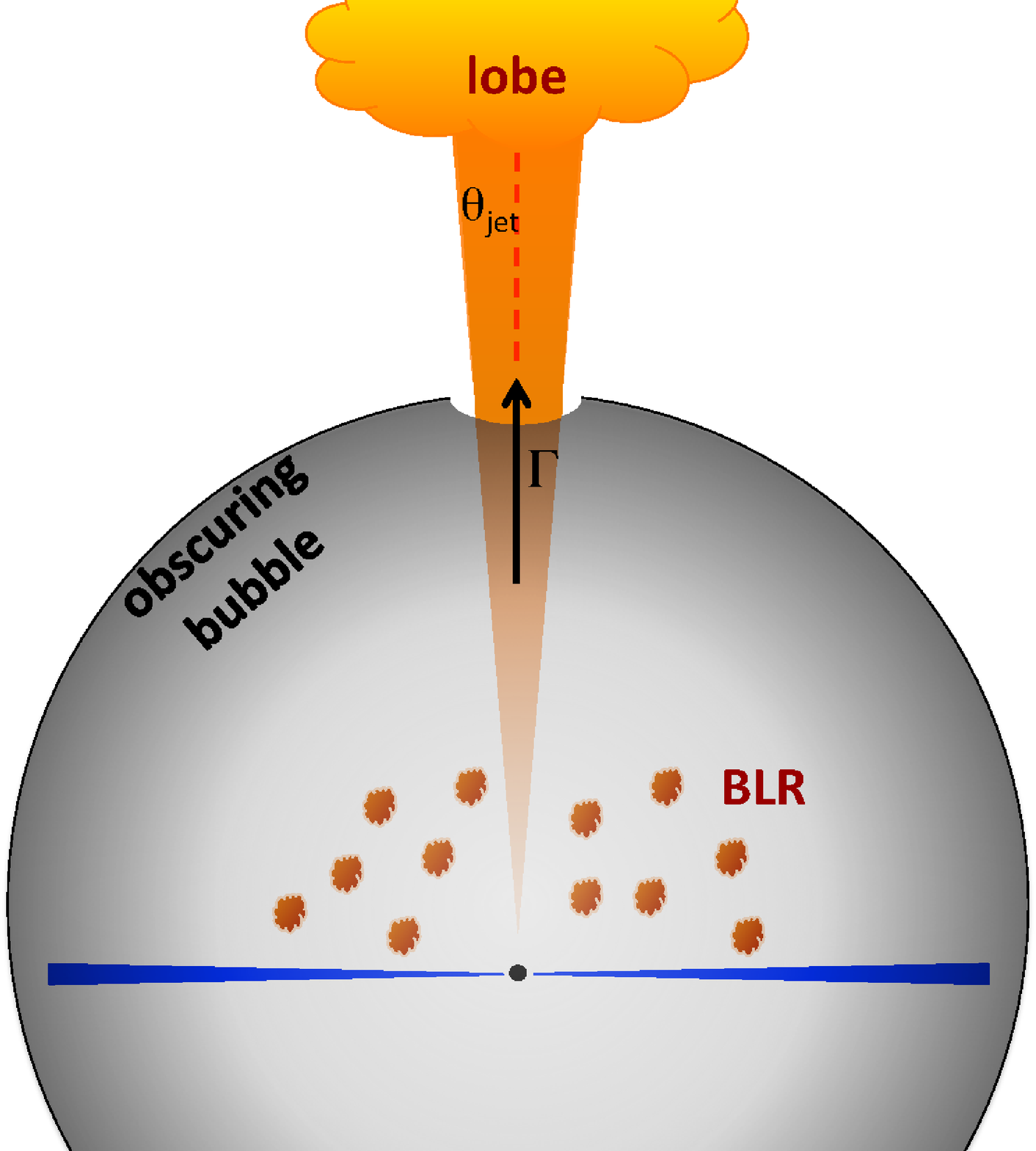,height=6cm} 
\caption{Cartoon of the proposed scenario (not to scale).
The grey sphere is the obscuring bubble, pierced by the
jet (in orange). The accretion disc (blue) and the BLR (brown) are
unobscured only if the source is observed down the jet.
} 
\label{bubbles}
\end{figure}

\section{Discussion}

Let us assume that the obscuring bubbles exist not only in jetted sources, but are
common to all high redshift quasars, including radio--quiet ones.
The evolution in time of the obscuring bubbles and the central black hole mass could however 
be different in jetted and non--jetted sources.

In fact, the presence of a jet could affect the accretion efficiency
$\eta_{\rm d}$, defined as $L_{\rm d}=\eta_{\rm d}\dot M c^2$: part of the dissipation of the 
gravitational energy could amplify the magnetic field instrumental to launch the jet.
In other words, while in the case of non--jetted AGN the gravitational energy is 
dissipated only through radiation from the disc (i.e.\ $\eta_{\rm d}=\eta$), radio--loud 
sources could use a fraction $f$ of the released gravitational energy to heat the disc,
and the remaining fraction $(1-f)$ to launch the jet (Jolley \& Kuncic 2008; Jolley et al. 2009):
\begin{equation}
\eta_{\rm d} \, =\, f\, \eta 
\end{equation}
This condition could lead to different evolution patterns of the obscuring bubbles. 
If we assume an Eddington--limited accretion  
until the obscuring bubble is blown away by the reached $L_{\rm th}$, 
the mass growth rate of the black hole is:
\begin{equation}
\dot M \, = \, \frac{dM}{dt} = \frac{1-\eta}{\eta_{\rm d}}\frac{L_{\rm Edd}}{c^2} 
			 = \frac{1-\eta}{\eta_{\rm d}}k M; \quad k={4\pi Gm_{\rm p} \over \sigma_{\rm T}c}
\end{equation}
where $m_{\rm p}$ is the proton mass, $\sigma_{\rm T}$ is the Thomson cross section and 
$G$ is the gravitational constant.
Therefore the black hole mass evolves as:
\begin{equation}
M(t; \eta_{\rm d}) = M_0\; {\rm exp}\left\{\frac{1-\eta}{\eta_{\rm d}} k\, t\right\}
\end{equation}
The threshold luminosity can therefore be expressed as a function of time:
\begin{eqnarray}
L_{\rm th} &=& 1.3\times10^{38} {M_{\rm th}(t; \eta_{\rm d})\over M_\odot}  \,\,\, {\rm erg \, s^{-1}}
   \nonumber \\
		&=& 1.3\times10^{38} {M_0\over M_\odot} \; {\rm exp}\left\{\frac{1-\eta}{\eta_{\rm d}} 
		k\, t_{\rm th}\right\}\,\,\, {\rm erg \, s^{-1}}
\end{eqnarray}
from which we can derive how much time it takes for a massive black hole 
to reach the threshold luminosity itself
\begin{equation}
t_{\rm th} = \frac{\eta_{\rm d}}{k(1-\eta)} 
\ln\left\{\frac{L_{\rm th}}{1.3\times10^{38}M_0/M_\odot}\right\} 
			\propto f\, {\eta\over 1-\eta}. 
\end{equation}
Considering the difference in the use of gravitational energy in
jetted and non--jetted AGN, there is a clear difference in the time needed for a source 
to blow away the dark bubble: if radio--loud AGN dissipate in radiation only $f=1/2$ of 
the released gravitational energy, 
{\it radio--loud AGN can get rid of their dark bubbles in half time, compared 
to non--jetted sources.}

\begin{figure} 
\vskip -0.3 cm
\psfig{file=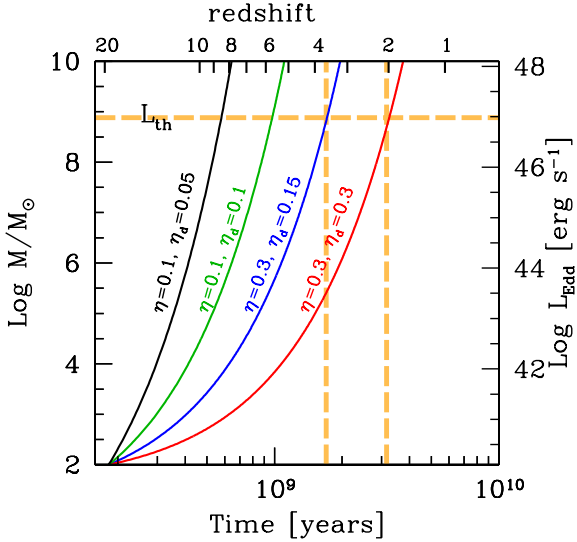,height=8cm} 
\vskip -0.4 cm
\caption{
Growth of a black hole mass for different values of the total efficiency $\eta$ and 
disc efficiency $\eta_{\rm d}$.
The horizontal dashed line corresponds to an assumed $L_{\rm th}=10^{47}$ erg s$^{-1}$.
The vertical dashed lines correspond to the times (bottom axis) and the redshift (top axis)
when the disc luminosity reaches $L_{\rm th}$. 
} 
\label{growth}
\end{figure}

On the other hand, the black hole mass, at the time $t_{\rm th}$, is independent of $f$.
For illustration, let us compare jetted and non--jetted sources of equal seed black hole mass $M_0$,
all emitting at their Eddington luminosity.
Jetted sources have black holes that grow faster (if $\eta_{\rm d}=f\eta$). 
Therefore, at any given time,  
their Eddington luminosity is larger than that of 
the radio--quiet ones accreting with the same total $\eta$, but with $\eta_{\rm d}=\eta$.
Fig. \ref{growth} shows the growth of the black hole for different values of $\eta$ and $\eta_{\rm d}$,
assuming that the accretion starts at $z=20$ on a seed black hole mass of 100 $M_\odot$.
Assuming a threshold luminosity of $L_{\rm th}=10^{47}$ erg s$^{-1}$, this is reached
first by the jetted sources.
Fig. \ref{growth} shows also the case of a total efficiency $\eta=0.1$.
Although we note the same trend (jetted sources with $\eta_{\rm d}=0.05$ grows faster), 
we can note that in this case the threshold luminosity $L_{\rm th}$ is reached at much
larger redshifts. 
At $z>4$, all jetted sources would have lost their absorbing bubble, 
and would be visible. 
One could also have jetted sources with $\eta=0.3$ (and a smaller $\eta_{\rm d}$), 
but radio--quiet sources with $\eta=\eta_{\rm d}\sim 0.1$.
In this case the radio--quiet ones could blow out the absorbing bubbles earlier than 
the jetted sources.
This does not affect the general picture we are proposing, but it seems unlikely that
at very early times, when we have large accretion rates, the spin of the black hole
(that controls the efficiency $\eta$) 
is less than its maximum value (Thorne 1974), for all kind of objects.
Major mergings could reset the black hole spin to values smaller than unity, but
the rarity of very large black hole masses and the short available time (of the order of 1 Gyr)
makes this possibility unlikely.

In the case we are discussing (all sources have $\eta=0.3$, but $\eta_{\rm d}$ of radio quiet is larger than
in radio--loud), we have an interesting consequence.
If we consider very large black hole masses (larger than $10^9 M_\odot)$, 
jetted sources becomes fully visible in the optical at earlier times
than radio--quiet objects.
Even if the {\it intrinsic} ratio $N_L/N_Q$ between the number of jetted and non jetted sources 
were constant in time (e.g. $N_L/N_Q=0.1$, as at low redshift), 
we would infer at $z\gsim$4 a radio--loud fraction larger than $N_L/N_Q$ from the blazar population.
We stress that this would be true only if we consider large black hole masses,
that need $\sim$ 1 Gyr to blow up the absorbing bubble.
If the critical luminosity $L_{\rm th}$ is smaller, it can be produced by a black hole of a smaller
mass, that is reached at earlier times (larger redshifts).
In this case, at $z\sim 4$, these sources are all visible, since they have already blown up their bubbles.

This dark bubble scenario makes a simple prediction: most high--$z$ parents of blazars with large
black hole masses should be absorbed in the optical band, but should be very bright in the infrared.
In this respect we can look at high--$z$ radio--galaxies.
Indeed, there is already one interesting example, 4C 41.17 ($z\sim$3.8), that is extremely bright
in the far infrared (with flux densities ranging from 23.4$\pm$2.4 $\mu$Jy at 3.6$\mu$ to 
36.5$\pm$3.5 $\mu$Jy at 8$\mu$ and   
a luminosity exceeding $10^{47}$ erg s$^{-1}$), but fainter in the
optical by a factor $\sim$30 
(Seymour et al. 2007;
Chambers et al. 1990;
van Breugel et al. 1998; 
Wu et al, in preparation).
This is not a proof of a $4\pi$ absorbing bubble, but suggests that the absorbing material
intercepts a larger fraction of the visible light, compared to local radio--galaxies.

%
%

\section*{Acknowledgements}
We thank the referee and Ann Wehrle for
their useful comments and
M. Lucchini for discussions.
We acknowledge financial contribution from the agreement ASI--INAF I/037/12/0
(NARO 15) and from the CaRiPLo Foundation and the
regional Government of Lombardia for the project ID 2014-1980 ``Science and technology
at the frontiers of $\gamma$--ray astronomy with imaging atmospheric Cherenkov Telescopes".



\end{document}